\documentstyle[12pt,epsf,epsfig]{article}
\begin{document}

\title{The Possible Interpretations of Y(4260)}
\author{Shi-Lin Zhu$^{1,2}$\\
$^1$Department of Physics, Peking University, Beijing 100871,
China\\
$^2$RCNP, Osaka University, Japan} \maketitle

\begin{abstract}

The recently observed Y(4260) lies far above the decay threshold
with a width less than 100 MeV. We argue that it's very difficult
to accommodate Y(4260) as a conventional $c\bar c$ radial
excitation or a D-wave state. It can't be a hadronic molecule. Its
production mechanism and special decay pattern do not favor the
glueball interpretation. If Y(4260) is a scalar tetraquark, it
must be produced by the $I=0$ component of the virtual photon.
Then the $I=1, I_z=0$ component of the virtual photon should have
produced its isovector partner $Y^\prime (4260)$, which may be
searched for in the decay channel $\pi^+\pi^-\pi^0 J/\psi$ using
exactly the same database from the initial state radiation
process. The observation/non-observation of $Y^\prime (4260)$ can
easily confirm/reject the tetraquark hypothesis. However, a
tetraquark far above threshold can fall apart into $D\bar D,
D^\ast\bar D$ very easily. Its not-so-large width and the
non-observation of $D\bar D$ mode tend to disfavor the tetraquark
hypothesis. Hence the only feasible interpretation is a hybrid
charmonium if Y(4260) is {\sl NOT} an experimental artifact. At
present, none of the experimental information from BABAR
measurement is in conflict with the hybrid charmonium picture.


\end{abstract}

\vspace{0.3cm}

\pagenumbering{arabic}

\section{Introduction}

The past several years have witnessed a revival of interest in
hadron spectroscopy due to the dramatic experimental progress.
Many new narrow resonances were observed, most of which lie close
to the threshold. Among them, we have $D_{sj}(2317)$
\cite{dsj2317}, $D_{sj}(2460)$ \cite{dsj2460}, $D_{sj}(2632)$
\cite{selex}, $\Theta^+$ \cite{leps}, X(3872) \cite{x3872},
Y(3940) \cite{y3940}.

Now the pentaquark legend is fading \cite{zsl1}. $D_{sj}(2632)$
was not confirmed by other groups. One may wonder whether it's an
experimental artifact \cite{zsl2}. Luckily $D_{sj}(2317),
D_{sj}(2460)$ and X(3872) were confirmed by other collaborations
and survived as well-established resonances. It's plausible that
$D_{sj}(2317)$ and $D_{sj}(2460)$ are the conventional $(0^+,
1^+)$ doublet in the heavy quark effective field theory while
their masses are lowered through their coupling to the nearby $DK$
and $D^\ast K$ thresholds \cite{dai}.

The hidden charm state X(3872) remains mysterious. According to
Ref. \cite{barnes}, the conventional $1^{++}$ quark model lies
several tens MeV higher than X(3872). In Ref. \cite{chrigg} the
authors used a potential model to include the coupled-channel
effects the open charm threshold which increases the "bare" $2
^3P_1$ state by 30 MeV. The previously prevailing molecule picture
\cite{swanson} encountered serious challenges recently. The
molecule picture predicted a tiny branching ratio for the decay
mode $X(3872) \to D\bar D \pi$, which is two orders of magnitude
smaller than the recent experimental data \cite{bauer,bernard}.
Moreover, the molecule picture predicted the ratio between $B^0\to
X(3872)K^0$ and $B^+\to X(3872)K^+$ to be much less than $10\%$
\cite{braaten} while the experimental data is order unity,
$0.61\pm 0.36\pm 0.06$ \cite{bernard}. The large branching ratio
of $D\bar D \pi$ mode indicates that the core of X(3872) is
probably the conventional $1^{++}$ quark model state, which is
only several tens of MeV higher than the $D{\bar D}^\ast, \rho
J/\psi, \omega J/\psi$ thresholds. There may exist strong coupling
between the bare quark model state and the continuum, which lowers
the $1^{++}$ bare state and transforms this state into X(3872). In
other words, X(3872) is a mixture of the dominant $1^{++}$ bare
$c\bar c$ state and a small portion of $D{\bar D}^\ast, \rho
J/\psi, \omega J/\psi$ continuum. Similar views can be found in
\cite{chao}. Phenomenology based on this new picture is still
lacking.

Very recently BABAR collaboration observed a broad resonance
around 4.26 GeV in the $\pi^+\pi^- J/\psi$ channel \cite{babar}.
Its decay width is around $(50\sim 90)$ MeV. A remarkable feature
of this state is that it lies far above threshold but still has a
rather narrow width.

Up to now, there are $125\pm 23$ total events \cite{babar}. Since
this resonance is found in the $e^+e^-$ annihilation through
initial state radiation $e^+e^-\to \gamma_{ISR} Y(4260)$, its
spin-parity is determined to be $J^{PC}=1^{--}$. However, this
state is not observed in the direct search $e^+e^-\to Y(4260)$
because of its low ratio $0.34\%$, buried by the $4\%$
experimental uncertainty for the hadronic cross section
\cite{babar}. The non-observation strongly indicates that
$Y(4260)\to \pi^+\pi^- J/\psi$ is one of the dominant decay modes.
At least the branching ratio of this mode should be much larger
than that of $D\bar D$. Otherwise the latter mode should have been
observed \cite{babar}.

\section{Possible Interpretations}

If Y(4260) {\sl really} exists, the available experimental data
contains useful information on its underlying structure. In fact,
BABAR may have found the elusive hybrid charmonium. We shall
analyze various possibilities one by one, assuming Y(4260) is a
genuine resonance and not an experimental artifact.

\subsection{Conventional $c\bar c$ State}

With $J^{PC}=1^{--}$, a conventional $c\bar c$ state is either a
radial excitation or a D wave state. From PDG \cite{pdg}, the
masses of the well-established radial excitations: $\psi(2S),
\psi(3S), \psi(4S)$ are 3686 MeV, 4040 MeV, 4415 MeV respectively.
The masses of $\psi(1 ^3D_1)$ and $\psi(2 ^3D_1)$ are 3770 MeV and
$(4159\pm 20)$ MeV \cite{pdg}. The $\psi(3 ^3D_1)$ state should
lie above 4.5 GeV. Therefore, it's nearly impossible to
accommodate Y(4260) as a conventional charmonium radial excitation
or D-wave state.

\subsection{Couple-Channel Effects}

One may wonder whether Y(4260) may arise from some couple-channel
effects. I.e., the bare charmonium state couples strongly with the
continuum. In this way, the bare quark model mass is shifted below
while its wave function admits some contribution from the
continuum as in the case of $D_{sj}(2317, 2460)$ and X(3872).

Among all the known cases where the couple-channel effect plays an
important role, there is a common feature. The resonance lies very
close to the decay threshold. A well-known example is the
$f_0/a_0(980)$, which is within 10 MeV of $K\bar K$ threshold. The
recent X(3873) sits nearly exactly on the threshold. $D_{sj}(2317,
2460)$ lies also very close to $DK, D^\ast K$ threshold. Quantum
mechanics tells us that the mixing strength is related to the
inverse of the energy difference of the two states.

Y(4260) is far away from both the open-charm and hidden charm
decay threshold. The analysis in Ref. \cite{chrigg} shows that the
mass shift of the $c\bar c$ state caused by the coupled-channel
effects of the open charm thresholds is only around tens MeV. If
the PDG charmonium assignments are correct, it's very difficult to
shift the mass of the $3 ^3D_1$ state from above 4.5 GeV down to
4.26 GeV by the coupled-channel effects.

\subsection{Molecule}

Hadronic molecules are states which lie below the continuum
threshold and carry a small amount of binding energy. Y(4260) lies
far above the $D\bar D, \omega J/\psi$ threshold.

One may wonder whether Y(4260) is a hadronic molecule composed of
${\bar D}_s D_{sj}(2317)$ since it is only 26 MeV below ${\bar
D}_s D_{sj}(2317)$ threshold. However the angular momentum and
parity conservation exclude the possibility of ${\bar D}_s
D_{sj}(2317)$ forming a $1^{--}$ molecular state.

Y(4260) is 24 MeV below $\bar D D_1(2420)$ threshold. $D_1(2420)$
is a very narrow resonance. Its total width of $(50\sim 90)$ MeV
disfavors the identification as a $\bar D D_1(2420)$ molecular
state. The total width also disfavors the identification as a
$\bar D D_1^\prime$ or $\bar D_0(2310) {\bar D}^\ast$ or ${\bar
D}^\ast D_1^\prime$ molecule since both $D_0(2310)$ and
$D_1^\prime$ are very broad resonances. Hence Y(4260) can not be a
hadronic molecule.

\subsection{Glueball}

C parity requires that a $J^{PC}=1^{--}$ glueball contains at
least three gluons. The decay patterns of the glueballs are
expected to be flavor-blind. If it is a glueball, Y(4260) should
decay into multiple light mesons very easily with so large phase
space. Certainly $\pi^+\pi^- J/\psi$ can not be a dominant decay
mode. On the other hand, Y(4260) is produced through initial state
radiation. The photon does not couple to gluons. Hence the
possibility of Y(4260) being a glueball is tiny.

\subsection{Tetraquark}

Naively one may speculate Y(4260) could be a tetraquark with the
quark content ${1\over \sqrt{2}} (u\bar u +d \bar d) c\bar c$. In
fact the presence of a heavy quark pair favors the formation of a
tetraquark since the heavy quark pair lower the kinetic energy
significantly.

With $J^{PC}=1^{--}$, one orbital excitation is required for a
tetraquark. Hence its mass is roughly
\begin{equation}
M_Y=2(m_c+ m_u)+ \Delta E_L
\end{equation}
where $m_c, m_u$ is the charm and up quark constituent mass,
$\Delta E_L$ is the orbital excitation energy. With $m_c=1.63$
GeV, $\Delta E_L=(0.30\sim 0.45)$ GeV \cite{barnes}, $m_u=0.33$
GeV, the mass of this tetraquark is roughly $(4.22\sim 4.37)$ GeV,
very close to the experimental value 4.26 GeV. The above
discussion shows the attractive side of the tetraquark hypothesis.

However, this hypothesis faces one serious obstacle: the
not-so-large decay width measured by BABAR Collaboration
\cite{babar}, $\Gamma_Y =(50\sim 90)$ MeV. If Y(4260) is a scalar
tetraquark with the quark content ${1\over \sqrt{2}} (u\bar u +d
\bar d) c\bar c$, it can easily decay into $D\bar D$ final states
via P-wave. Such a decay occurs through the so-called
"super-allowed" "fall-apart" mechanism, which is simply the
regrouping of the two quarks and anti-quarks inside the tetraquark
into two color-singlet mesons. No symmetry forbids the decay of
$Y(4260)\to D\bar D$. With enough phase space, one would expect
its total decay width to be around several hundred MeV or more,
contrary to the experimental value $(50\sim 90)$ MeV. Moreover,
$D\bar D$ should be one of the dominant decay modes, which is also
against BABAR's observation.

In fact, BABAR Collaboration can further confirm/reject the
tetraquark conjecture through the observation/non-observation of
the isovector tetraquark $Y^\prime (4260)$ in the $\pi^+\pi^-\pi^0
J/\psi$ channel by analyzing the same set of initial state
radiation data.

If Y(4260) is a tetraquark, it must belong to a $SU_F(3)$ flavor
multiplet. In the present case, we have a nonet from $3_F\times
\bar 3_F= 1_F+8_F$. In the vector channel, the flavor eigenstates
always split into the mass eigenstates because of the ideal
mixing. A typical example is that the physical states are $\omega$
and $\phi$ instead of $\omega_0$ and $\omega_8$. We list the quark
content of these nine states below: (1) I=0, Y(4260), ${1\over
\sqrt{2}} (u\bar u +d \bar d) c\bar c$; (2) I=0, $Y_{ss}(4560)$,
$s\bar s c\bar c$; (3) $I=1, I_z=0$, $Y^\prime(4260)$, ${1\over
\sqrt{2}} (u\bar u -d \bar d) c\bar c$; (4) $I=1, I_z=\pm 1$,
$Y^{\prime \pm}(4260)$, $u\bar d c\bar c$, $d\bar u c\bar c$; (5)
$I={1\over 2}, I_z=\pm {1\over 2}, Y_s^\prime (4410), u\bar s
c\bar c$, $d\bar s c\bar c, s\bar d c\bar c, s\bar u c\bar c$.

The neutral isovector state $Y^\prime (4260)$ deserves special
attention. With $J^{PC}=1^{--}$ and $I=1$, its G-parity is
positive. While $J/\psi$ carries negative G-parity, the
conservation of G-parity demands $Y^\prime (4260)$ decays into
three or five or seven pions.

The isospin of the virtual photon is indefinite. It contains both
$I=1, I_z=0$ and $I=0, I_z=0$ components. In the extreme case that
Y(4260) really turns out to be an isoscalar tetraquark, it must
have been produced by the $I=0, I_z=0$ component of the virtual
photon in the $e^+e^-$ annihilation through initial state
radiation. Then the accompanying isovector tetraquark $Y^\prime
(4260)$ should also have been produced by the $I=1, I_z=0$
component of the virtual photon with the same ratio. BABAR
Collaboration may search the $\pi^+\pi^-\pi^0 J/\psi$ channel to
look for the possible $Y^\prime (4260)$ signal in their initial
state radiation data.

Moreover we urge BABAR Collaboration to scan their data to search
for the possible $Y_{ss}(4560)$ signal in the $K\bar K J/\psi$
final states. It will also be useful to study the angular
distribution to find out whether the decay proceeds via S-wave.

In short, more experimental investigation is needed to completely
exclude the possibility of $Y(4260)$ being a tetraquark although
the available data already disfavors the tetraquark hypothesis.

\subsection{Hybrid Charmonium}

Naively when one gluon is confined within a hadron bag, its
binding energy is roughly around 1.1 GeV. Hence the
$J^{PC}=1^{--}$ hybrid charmonium mass may be around $4.2$ GeV.
Some lattice QCD calculation found that the exotic $1^{-+}$ hybrid
charmonium lies around 4.3 GeV \cite{lattice2,lattice3}. There is
no lattice QCD calculation of $1^{--}$ hybrid charmonium mass.

According to QCD sum rule calculation \cite{zsl3} and the analysis
in the flux tube model \cite{page}, a hybrid meson tends to decay
into final states with one L=1 meson. An extensive study of open
charm decay modes can be found in Ref. \cite{flux}. It is very
interesting to note that these author did predict a small $D\bar
D$ width and a rather narrow $1^{--}$ hybrid charmonium if its
mass is below 4.3 GeV. BABAR may search the other decay modes as
suggested in \cite{flux}.

There is not much discussion of the hidden charm decay modes of
the hybrid charmonium in literature. Physically these decays may
occur via the following process: $c\bar c G\to J/\psi + GG \to
J/\psi + \pi^+\pi^-$. Because of the isospin symmetry, Y(4260)
also decays into $\pi^0\pi^0 J/\psi, \eta \eta J/\psi$. Motivated
by the discovery mode, we suggest the decay modes $Y(4260)\to
\omega + \chi_{c0, c1, c2}\to 3\pi + \chi_{c0, c1, c2}$ may be
important if Y(4260) is really a hybrid meson.

\section{Discussion}

We have excluded the possibility of Y(4260) being a conventional
$c\bar c$ state, a molecule, a glueball one by one using the
available experimental information. The tetraquark hypothesis is
not favored by its small total width and non-observation of $D\bar
D$ decay mode. In order to completely exclude this possibility,
BABAR may search for its isovector partner using the same database
in the $\pi^+\pi^-\pi^0 J/\psi$ channel. The most plausible
interpretation is a hybrid charmonium, which is consistent with
all the experimental information.

No matter whether Y(4260) is a tetraquark state or hybrid
charmonium, it shall decay into $\pi^0\pi^0 J/\psi$ if it's a
genuine resonance. This branching ratio is $25\%$ of the discovery
mode from isospin symmetry consideration. The other possible mode
is $\eta \eta J/\psi$, which may be suppressed by phase space. The
electromagnetic decay modes $Y(4260)\to \gamma + \chi_{c0, c1,
c2}$ are also very interesting. Confirmation of Y(4260) in a
different channel and by a different experimental group in the
near future will greatly shed light on the charmonium
spectroscopy.

\section{Acknowledgment}

This project was supported by the National Natural Science
Foundation of China under Grants 10375003 and 10421003, Ministry
of Education of China, FANEDD, Key Grant Project of Ministry of
Education (NO 305001) and SRF for ROCS, SEM.


\end{document}